\pdfoutput=1

\documentclass[12pt,review]{elsarticle}
\usepackage[margin=2.5cm]{geometry}

\usepackage{graphicx}
\usepackage{amssymb}
\usepackage{tabularx}
\usepackage{booktabs} 
\usepackage{siunitx}
\usepackage{units}
\usepackage{placeins}
\usepackage{color}
\usepackage{soul}
\usepackage{textcomp}
\usepackage{hyperref}
\usepackage{lineno}
\usepackage{multirow}

\hypersetup{
	colorlinks   = true, 
	urlcolor     = blue, 
	linkcolor    = blue, 
	citecolor   = red 
}

\biboptions{sort&compress}

\journal{Composites Science and Technology }

\begin{document}
	
\begin{frontmatter}

\title{The evolution of the structure and mechanical properties of fully bioresorbable polymer-glass composites during degradation}

\author[label1]{Reece N. Oosterbeek \corref{cor1}}
\ead{rno23@cantab.ac.uk; r.oosterbeek@imperial.ac.uk}

\author[label2]{Xiang C. Zhang}
\author[label1]{Serena M. Best}

\author[label1]{Ruth E. Cameron  \corref{cor1}}
\ead{rec11@cam.ac.uk}

\cortext[cor1]{Corresponding author}

\address[label1]{Cambridge Centre for Medical Materials, Department of Materials Science and Metallurgy, University of Cambridge, Cambridge, United Kingdom}
\address[label2]{Lucideon Ltd, Queens Road, Penkhull, Stoke-on-Trent, United Kingdom}

\begin{abstract}

Fully bioresorbable polymer matrix composites have long been considered as potential orthopaedic implant materials, however their combination of mechanical strength, stiffness, ductility and bioresorbability is also attractive for cardiac stent applications. This work investigated reinforcement of polylactide-based polymers with phosphate glasses, addressing key drawbacks of current polymer stents, and examined the often-neglected evolution of structure and mechanical properties during degradation. Incorporation of 15 – 30wt.\% phosphate glass led to modulus increases of up to 80\% under simulated body conditions, and 15wt.\% glass composites retained comparable ductility to pure polymers, crucial for stent applications where ductility and stiffness are required. Two-stage degradation was observed, dominated by interfacial water absorption and glass dissolution. Polymer embrittlement mechanisms (crystallisation, enthalpy relaxation) were suppressed by glass addition, allowing composites to achieve a more controlled loss of mechanical properties during degradation, which could allow gradual transfer of loading to newly healed tissue. These results provide a valuable new system for understanding the structural and mechanical changes occurring during degradation of fully bioresorbable polymer matrix composites, providing important new data to underpin the design of effective cardiac stent materials.\\

\end{abstract}


\end{frontmatter}


\section{Introduction}

Bioresorbable stents have the potential to revolutionise interventional cardiology, reducing rates of restenosis, late thrombosis, and other long-term complications \cite{Iqbal2014}. Unfortunately this promise has not yet been realised, and many current limitations arise from the material mechanical properties. PLLA (poly-L-lactide) is a widely used stent material due to its bioresorbability and superior mechanical strength compared with other bioresorbable polymers, however it still has a significantly lower yield strength than the metals used for stents. This necessitates thicker struts to provide sufficient device strength, increasing blood flow turbulence and inhibiting stent endothelialisation, increasing the risk of thrombosis \cite{McMahon2018}. The relatively low elastic modulus of PLLA is also an issue when considering the elastic recoil of a stent after implantation, which must be minimised to prevent restenosis or stent dislodgement \cite{McMahon2018}. It is clear that to overcome issues with existing bioresorbable stent technologies, improvements in material yield strength and elastic modulus are necessary \cite{Mukherjee2017}.

Polymer matrix composites have been widely investigated for use as orthopaedic implants \cite{Gohil2017} due to the ability to improve mechanical properties by incorporating components such as hydroxyapatite or $\alpha$-TCP (tricalcium phosphate) \cite{Naik2017, Wilberforce2011}. Composites incorporating bioactive glasses are also attractive due to the exceptional tunability of the dissolution rate and ion release properties that can be achieved by engineering the glass composition \cite{Knowles2003, Boccaccini2010, Felfel2013}. For applications as bioresorbable stent materials however, these composites have not been frequently studied.

Typically the elastic modulus is improved by addition of inorganic particles to a polymer matrix, due to the higher stiffness of the inorganic phase. The upper and lower bounds of the elastic modulus (Voigt-Reuss bounds) can be calculated, however a more detailed prediction for particulate composites is given by the Counto model \cite{Counto1964}:

\begin{equation}
	\label{eq:Counto}
	\frac{1}{E_{Counto}} = \frac{1-\sqrt{\phi_{f}}}{E_{m}} + \frac{1}{ \left( \frac{1-\sqrt{\phi_{f}}}{\sqrt{\phi_{f}}} \right)E_{m} + E_{f}}
\end{equation}

where subscripts $f$ and $m$ denote the filler and polymer matrix respectively, and $\phi_{f}$ is the filler volume fraction. The composite yield strength is also heavily influenced by the addition of filler particles to the polymer matrix, where interfacial adhesion and subsequent stress transfer between the matrix and filler is a key consideration. Dispersed particles can also hinder crack propagation, again improving the yield strength, but must be balanced against the effect of stress concentration in the polymer matrix around the filler particle \cite{Fu2008}. The upper and lower bounds of the composite yield strength $\sigma_{y,c}$ were therefore determined by Nicolais and Narkis \cite{Nicolais1971} as a function of the polymer matrix yield strength $\sigma_{y,m}$ and filler volume fraction $\phi_{f}$:

\begin{equation}
	\label{eq:NNcomp-upper}
	\sigma^{upper}_{y,c} = \sigma_{y,m}
\end{equation}
\begin{equation}
	\label{eq:NNcomp-lower}
	\sigma^{lower}_{y,c} = \sigma_{y,m}(1 - 1.21\phi_{f}^{\nicefrac{2}{3}})
\end{equation}

In a ductile matrix, the addition of brittle filler materials can significantly increase the brittleness of the composite due to stress concentration. However, if interfacial adhesion is good, ductility can be maintained or increased as a result of toughening mechanisms such as crack deflection and interfacial debonding, all of which increase the energy required for failure \cite{Fu2008}. To date most strategies for tuning the mechanical behaviour of stent materials rely on altering the polymer chemistry (i.e. copolymerisation ratio etc.) or polymer structure (orientation, crystallisation), however these methods cannot be applied without altering the degradation behaviour of the material \cite{McMahon2018, Han2009}. By incorporating bioactive glass however, results have shown that composite degradation can be tuned by varying the glass composition \cite{Kim2005, Stamboulis2002}. This occurs via several mechanisms. Increased water absorption at the polymer/glass interface, especially for continuous glass fibres where wicking plays a large role, can accelerate the degradation rate \cite{Ahmed2011}. There are also chemical effects resulting from dissolution of the inorganic filler. $\alpha$-TCP particles are seen to slow polymer degradation due to the buffering effect of the ions released as the ceramic dissolves \cite{Naik2017}. Similar effects can occur in polymer-glass composites, where ion exchange can result in a buffering effect at the interface \cite{Boccaccini2003}.

These phenomena lead to the idea that polymer-glass composites can allow decoupling of mechanical and degradation properties, whereby incorporation of glass particles can provide mechanical reinforcement, and changing the glass composition can tune the degradation profile, with minimal impact on the mechanical properties. This degradation profile is another key issue with current stents based on PLLA that must be addressed. PLLA is well known to experience very slow degradation \textit{in vivo}, requiring several years for full resorption, while healing after stent deployment is typically complete after around six months \cite{McMahon2018}. Developing materials with accelerated rates of resorption would therefore be highly advantageous for cardiac stent devices \cite{Mukherjee2017}.

Material mechanical properties also evolve significantly during degradation, as a result of structural changes. For bioresorbable polyesters these structural changes have been well characterised, with enthalpy relaxation and crystallisation providing temporary increases in strength and stiffness, while molecular weight reduction causes loss of strength, stiffness, and ductility over the long term \cite{Karjalainen1996, Han2009, Oosterbeek2019}. However the changes in structure and mechanical properties of composite materials during degradation, and the effect of filler on these polymer structural changes has not been extensively studied. This work aims to assess the effect of composite matrix composition, filler composition, and filler amount, on the degradation behaviour and development of mechanical properties during degradation for polymer-glass composites. Polymer matrix compositions of pure PLLA and 90PLLA:10PLCL(70:30)-PEG  (polyethylene glycol functionalised poly(\textsc{L}-lactide-co-$\varepsilon$-caprolactone)) were chosen to observe the effect of blend composition on composite properties. PLLA is used as a control and majority component to provide sufficient mechanical properties and bioresorbability, with a small amount of PLCL-PEG added to improve ductility and accelerate degradation, without causing rapid loss of mechanical integrity. This work shows for the first time how various structural phenomena (such as water absorption, polymer relaxation, crystallisation and degradation, glass dissolution and inorganic crystal deposition) interact and develop during polymer-glass composite degradation, and the resulting effect on composite properties.


\section{Materials and Methods}

\subsection{Materials and processing}
Phosphate glasses, with nominal composition (P\textsubscript{2}O\textsubscript{5})\textsubscript{90-\textit{x}}(CaO)\textsubscript{\textit{x}}(Na\textsubscript{2}O)\textsubscript{10}, where $x = 45$ or $50$, (denoted P45Ca45 and P40Ca50 respectively) were produced using a melt-quenching method described elsewhere \cite{Oosterbeek2021}. PLLA (Ingeo 2500 HP) was supplied by Natureworks LLC, USA, and PLCL-PEG was synthesised and supplied by Ashland Specialties Ireland Ltd. (Dublin, Ireland), with copolymer ratio of 70:30 (LA:CL), and a singular 550 g mol\textsuperscript{-1} PEG end-group. Polymer-glass composites were fabricated using a precipitation method described elsewhere \cite{Oosterbeek2020, Oosterbeek2021a}. Polymers (pure PLLA or a 90:10 w/w PLLA/PLCL-PEG blend) were dissolved in DCM, and glass slurry with particle size $d_{0.5}$ = 1.6 \textmu m  was added to give the desired polymer/glass ratio ($\phi_{f}$ = 0, 15, or 30 wt.\% glass). After mixing and sonicating, ethanol was added to produce solid composite precipitate, which was dried under vacuum at 50\textdegree C for 10 days to remove residual solvent. Composite precipitate was processed into dumbbell or disc-shaped samples using micro-injection moulding (IM 5.5, Xplore Instruments BV, The Netherlands) at the minimum temperature required for complete mould filling (243-273\textdegree C), with the custom-made mould held at ambient temperature.

\subsection{Characterisation}
Differential scanning calorimetry (DSC) was carried out using a DSC Q2000 (TA Instruments, USA), in hermetic Al pans at 20\textdegree C min\textsuperscript{-1}, from -20 to 230\textdegree C. TA Universal Analysis software was used to determine the glass transition temperature T\textsubscript{g}. X-ray diffraction (XRD) was carried out using a Bruker D8 Advance diffractometer with Cu K$\alpha$ radiation in a 2$\theta$ range of 5-50\textdegree, 0.05\textdegree\ step size, and 1.0 s step\textsuperscript{-1} dwell time. Line profile analysis was used to quantify the amorphous and crystalline content using Match! (Crystal Impact GbR) software, and unknown phases were identified using the Crystallography Open Database \cite{Grazulis2009}. Scanning Electron Microscopy (SEM) was carried out using an FEI Nova NanoSEM, using an accelerating voltage of 5 kV. Cross-sections were prepared by cryo-fracturing in liquid nitrogen and sputter coating with 10 - 20 nm Au. Degraded samples were vacuum dried at room temperature until reaching constant mass before DSC, XRD, or SEM. To determine the wet mass during degradation, samples were removed from solution, dabbed dry, and weighed using a Sartorius Ultramicro balance. To find the dry mass and water uptake samples were vacuum dried at room temperature until reaching constant mass. Ashing was then carried out at 650\textcelsius\ to burn off the polymer and determine the polymer weight fraction and fraction of glass particles remaining.

\subsection{Mechanical testing}
Tensile testing was carried out using a 1ST Benchtop Tester (Tinius Olsen Ltd, UK) with 1 kN load cell, under a constant elongation rate of 1 mm min\textsuperscript{-1}. Dumbbell samples (5 mm gauge length, 0.6 mm thickness) were tested in ambient (dry at 25\textdegree C) and simulated body conditions (immersed in deionised water at 37\textdegree C) using a Saline Test Tank with Heater (Tinius Olsen Ltd, UK). Strain was measured using a video extensometer and custom-built LabVIEW software. Yield strength ($\sigma_{y}$) was taken as the 0.2\% offset yield point.

\subsection{Degradation}
Degradation studies were carried out by immersing individual samples in PBS (phosphate-buffered saline, 0.3 mL per mg sample) at 37\textcelsius . pH measurements were taken regularly using an HI 4222 pH meter (Hanna Instruments Ltd., UK), and Ca\textsuperscript{2+} ion concentration was measured using an ISE (ion selective electrode - Sentek 361-75), calibrated using a modified Nernst equation \cite{Midgley1977} with standard solutions diluted with deionised water from a 0.1 mol L\textsuperscript{-1} calcium standard (Hanna HI 4004-01). PBS alone was used as a control for pH and Ca\textsuperscript{2+} ion measurements.


\section{Results}

\subsection{Initial structure and properties}

\begin{figure}[htb]
	\centering
	\includegraphics[width=8cm]{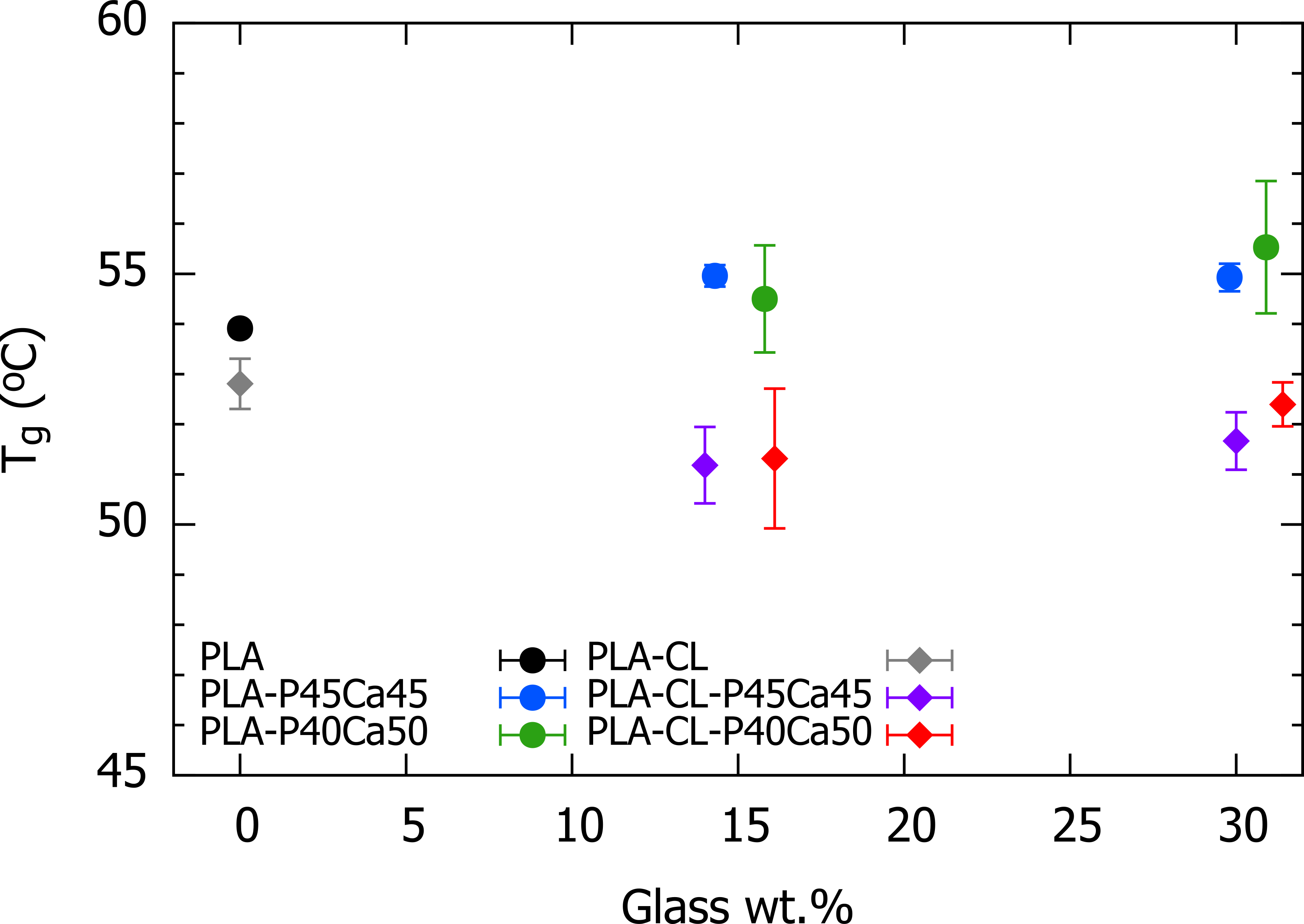}
	\caption{Glass transition temperatures ($T_{g}$) of various as-fabricated polymer-glass composite samples, according to measured glass content (wt.\%).}
	\label{fig:Comp_Tg_t0}
\end{figure}

Ash testing results (Table \ref{tab:compcomps}) showed complete polymer burn-off and demonstrated effective incorporation of the desired amount of filler. No sharp XRD peaks were observed for as-fabricated samples, indicating that both the polymer and glass components were in an amorphous state. Incorporation of glass into the composite was not seen to affect the $T_{g}$  in the as-fabricated state (Fig. \ref{fig:Comp_Tg_t0}) due to the relatively weak interactions between the polymer and phosphate glass, and the micron-scale glass particle size \cite{Wilberforce2011}, however the addition of PLCL(70:30)-PEG to PLLA reduced the $T_{g}$. Only one polymer $T_{g}$ was detected indicating that this addition of 10wt.\% PLCL(70:30)-PEG to PLLA formed a single phase blend as expected \cite{Oosterbeek2019}.

\begin{table}[htb]
	\centering
	\textbf{\caption{\label{tab:compcomps}Composites produced, showing weight fractions of different polymers within the matrix component, and the amount of glass filler - both nominal and measured by ashing.}}
	{\renewcommand{\arraystretch}{0.8}
		\begin{tabularx}{\columnwidth}{ p{4.1cm} l l l l p{2.7cm} }
			\hline
			Sample code & \multicolumn{2}{p{3.7cm}}{Matrix (wt.\% in matrix)} & \multicolumn{2}{p{2.9cm}}{$\phi_{f}$ (filler wt.\% of total)} & \multirow{2}{=}{Measured ash (wt.\% of total)} \\
			\cmidrule(r){2-3} \cmidrule(r){4-5}
			& PLLA & PLCL(70:30)-PEG & P45Ca45 & P40Ca50 &\\
			\hline
			PLA & 100 & - & - & - & 0.013 $\pm$0.007\\
			PLA-CL & 90 & 10 & - & - & 0.01 $\pm$0.01 \\
			PLA-0.15P45Ca45 & 100 & - & 15 & - & 14.3 $\pm$0.3 \\
			PLA-CL-0.15P45Ca45 & 90 & 10 & 15 & - & 14.0 $\pm$0.2 \\
			PLA-0.3P45Ca45 & 100 & - & 30 & - & 29.8 $\pm$0.1 \\
			PLA-CL-0.3P45Ca45 & 90 & 10 & 30 & - & 30.0 $\pm$0.1 \\
			PLA-0.15P40Ca50 & 100 & - & - & 15 & 15.8 $\pm$0.2 \\
			PLA-CL-0.15P40Ca50 & 90 & 10 & - & 15 & 16.1 $\pm$0.2 \\
			PLA-0.3P40Ca50 & 100 & - & - & 30 & 30.9 $\pm$0.1 \\
			PLA-CL-0.3P40Ca50 & 90 & 10 & - & 30 & 31.4 $\pm$0.1 \\
			\hline
		\end{tabularx}
	}
\end{table}

\begin{figure}[htb]
	\centering
	\includegraphics[width=13cm]{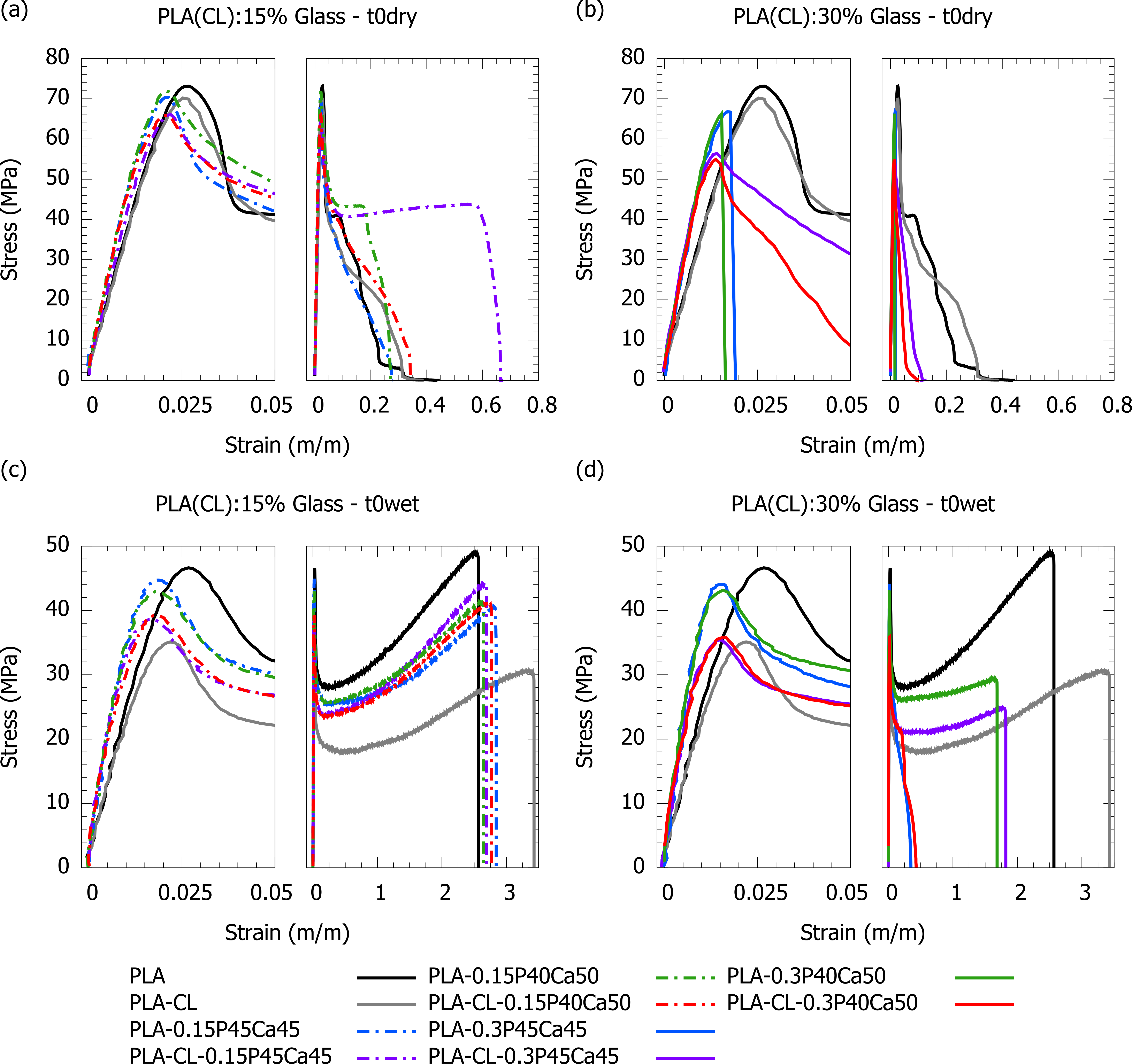}
	\caption{Example stress-strain curves for tensile testing of polymer-glass composites under various conditions. (a, b): As-fabricated composites with 15wt.\% glass (a) and 30wt.\% glass (b) tested dry at room temperature (\textbf{t0dry}). (c, d): As-fabricated composites with 15wt.\% glass (c) and 30wt.\% glass (d) tested immersed in 37\textcelsius\ water (\textbf{t0wet}).}
	\label{fig:compmech_ss_t0dry-wet}
\end{figure}

\begin{figure}[htb]
	\centering
	\includegraphics[width=13cm]{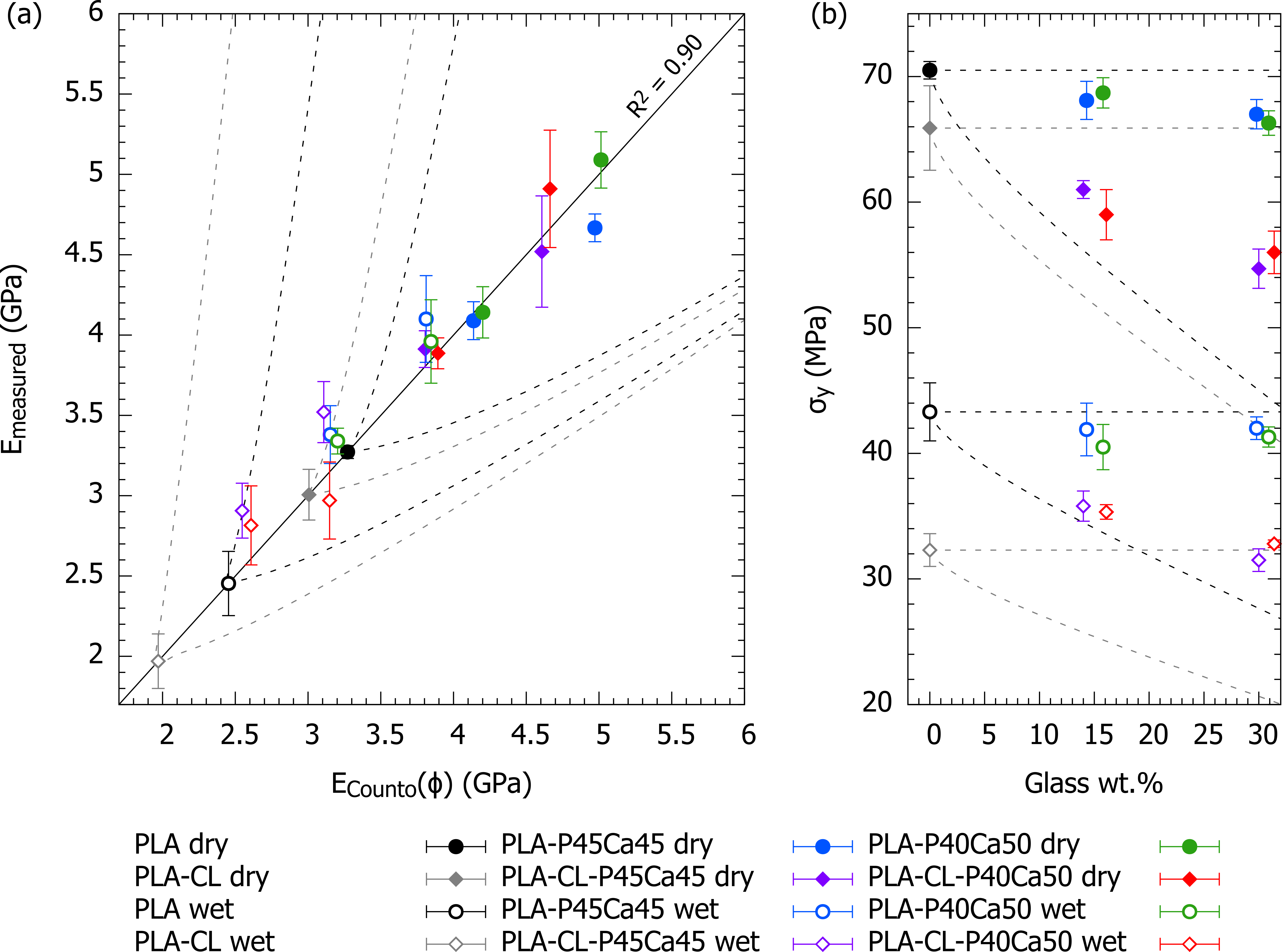}
	\caption{(a) Measured elastic modulus vs. elastic modulus calculated from the matrix properties, glass properties, and loading fraction, using the Counto model \cite{Counto1964}, with $E_{glass}$ = 48 GPa \cite{Sglavo2019}. Data is shown for composites based on PLA and PLA-CL polymer matrices, in wet (at 37\textcelsius ) and dry conditions. Dashed lines shown are the upper and lower bounds for the modulus (Voigt-Ruess bounds), and the solid line represents $E_{measured} = E_{Counto}$. (b) Yield strength for polymer composites tested dry at room temperature (dry), and immersed in 37\textcelsius\ water (wet). Dashed lines show the upper and lower bounds for the yield strength, according to Nicolais and Narkis \cite{Nicolais1971}.}
	\label{fig:E-yield-t0dry-wet}
\end{figure}

Representative stress-strain curves, and a summary of the mechanical properties of the composites produced are shown in Figs. \ref{fig:compmech_ss_t0dry-wet} and \ref{fig:E-yield-t0dry-wet}. Under dry conditions, the addition of phosphate glass was seen to provide significant mechanical reinforcement, increasing the elastic modulus, however the yield strength showed a modest decrease. The polymers, and composites with lower (15wt.\%) glass content showed similar deformation behaviour, with a small amount of plastic deformation after yield, however when the glass content was increased to 30wt.\% the ductility decreased noticeably, with failure occurring shortly after yield.

The addition of PLCL(70:30)-PEG to PLLA in the polymer matrix component resulted in a decrease in the elastic modulus across all compositions, as well as a reduction in yield strength, as expected from previous results \cite{Oosterbeek2019}. The addition of PLCL(70:30)-PEG to the polymer matrix also appeared to increase the elongation at break slightly, but did not lead to a large scale change in the deformation behaviour. The glass composition (P45Ca45 or P40Ca50) did not impact the composite mechanical properties under ambient conditions.

Significant changes were seen in these mechanical properties when tested in simulated body conditions (immersed in deionised water at 37\textcelsius ). For all compositions the elastic modulus significantly reduced, although the addition of phosphate glass still resulted in an increased elastic modulus. The yield strength also decreased significantly under simulated body conditions, while the slight reduction in yield strength observed in ambient conditions when glass was incorporated was no longer present under simulated body conditions. The elongation at break observed under simulated body conditions was significantly higher than when tested in ambient conditions. Large scale plastic deformation was seen for the polymer-only samples and composites with 15wt.\% glass, where at least 200\% elongation was observed before failure. Composites with 30wt.\% glass were also  more ductile than in ambient conditions, however the large scale plastic deformation seen for 15wt.\% glass composites was not consistently achieved, and the elongation at break was lower and more variable. 

\subsection{Long-term degradation behaviour}

\begin{figure}[h!]
	\centering
	\includegraphics[width=12cm]{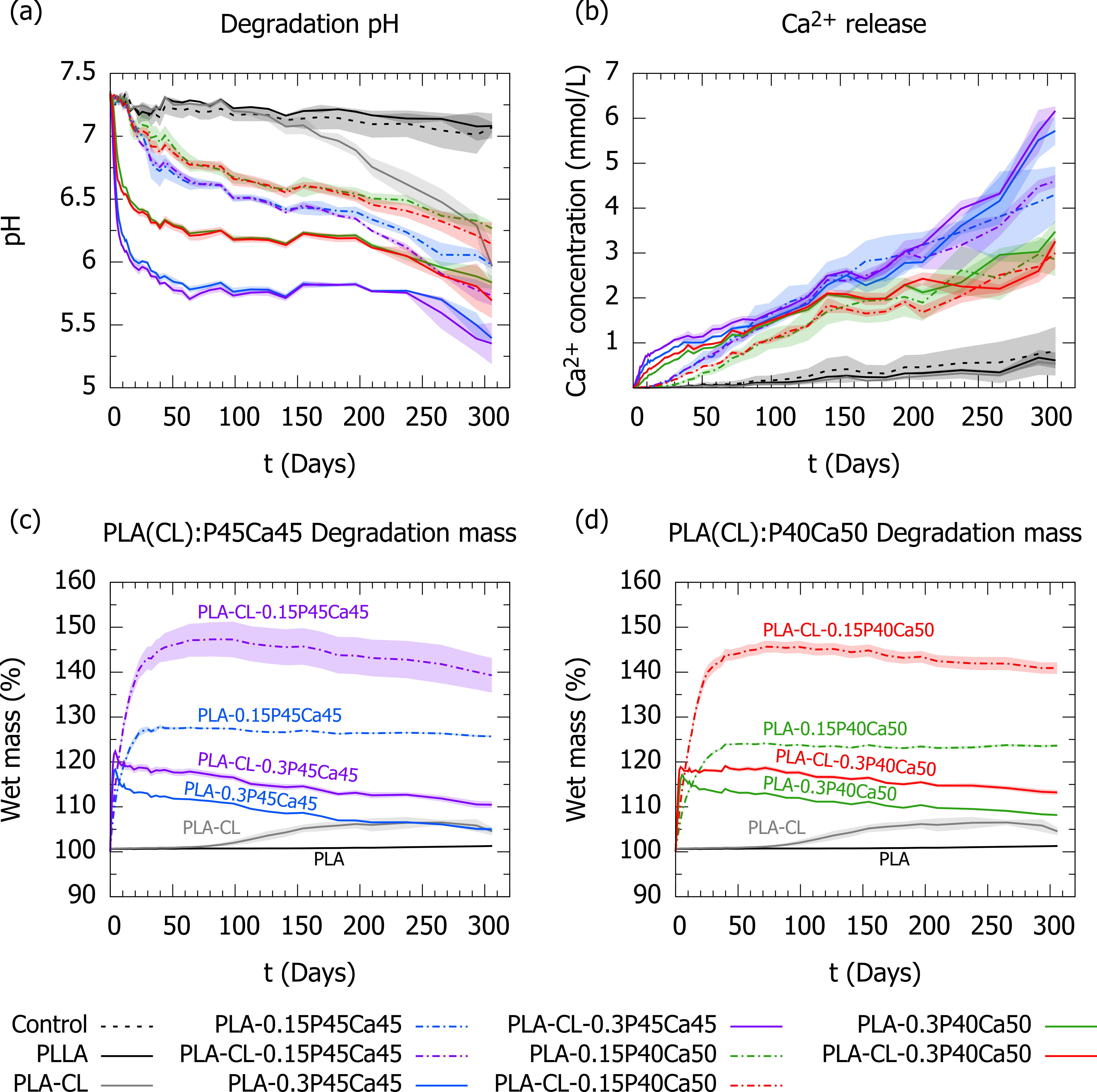}
	\caption{Degradation behaviour of composite samples in PBS at 37\textcelsius. Graphs show solution pH (a), solution Ca\textsuperscript{2+} concentration (b), and wet mass (\% of original mass) for composites with P45Ca45 glass (c) and P40Ca50 glass (d). Shaded regions denote standard deviation, n = 3.}
	\label{fig:Deg_data_all}
\end{figure}

Results from long-term degradation tests are shown in Fig. \ref{fig:Deg_data_all}. Due to the slow degradation of PLLA no pH change was seen for this composition, while the behaviour of PLA-CL was consistent with results on this same composition (90PLLA:10PLCL(70:30)-PEG) from \cite{Oosterbeek2019}, the onset of degradation after about 300 days. Significant differences were seen for the glass-containing composites, demonstrating a fast pH reduction before plateauing, as a result of glass dissolution. The plateau pH was dependent on the phosphate glass composition, with the higher P\textsubscript{2}O\textsubscript{5} glass resulting in a lower pH, consistent with previous results \cite{Oosterbeek2021}. Although the addition of PLCL-PEG had a major effect on the degradation of the unfilled matrix, the same cannot be said of the composite materials. For most of the degradation experiment no differences were observed based on the matrix composition, however after around 200 days composites with low glass content (15wt.\%) and a matrix containing PLCL-PEG did appear to show additional pH reduction compared with the equivalent composite without PLCL-PEG, however this difference was relatively small.

The slow increase in measured Ca\textsuperscript{2+} content observed for the control solution, PLA and PLA-CL samples can be attributed to contamination of the solution with residual ions from the electrodes used for pH and Ca\textsuperscript{2+} ion measurement. In spite of thorough washing between samples, some transfer and contamination was unavoidable. Composites containing glasses with higher P\textsubscript{2}O\textsubscript{5} content displayed higher Ca\textsuperscript{2+} release despite their lower CaO content, due to their faster dissolution. Composites with higher glass content also displayed higher Ca\textsuperscript{2+} release initially, however over time this difference disappeared, suggesting that some equilibrium was reached between the dissolving glass and the solution. Few differences were observed between composites with the PLA and PLA-CL matrices.

The pure polymers displayed low water absorption of around 1\% initially, with the PLA-CL sample later showing an increase of about 6\%, due to the increased water absorption that precedes bulk degradation \cite{Oosterbeek2019}. Composite samples demonstrated significantly higher wet mass during degradation, and the presence of PLCL-PEG in the matrix dramatically increased this in all cases. The most significant factor determining the wet mass increase however, appeared to be the glass content. Composites with 30wt.\% glass displayed a fast initial mass increase, followed by a peak and slow decrease, which can be attributed to fast water absorption and subsequent glass dissolution. Composites with 15wt.\% glass demonstrated similar behaviour, however the initial absorption was slower but reached a higher mass. The initial absorption stage did not appear to be heavily influenced by the glass composition, however in the later mass loss stage composites containing the faster dissolving P45Ca45 glass \cite{Oosterbeek2021} lost mass more quickly.

\subsection{Structure changes during degradation}

\subsubsection{Composite components}

\begin{figure}[htb]
	\centering
	\includegraphics[width=15cm]{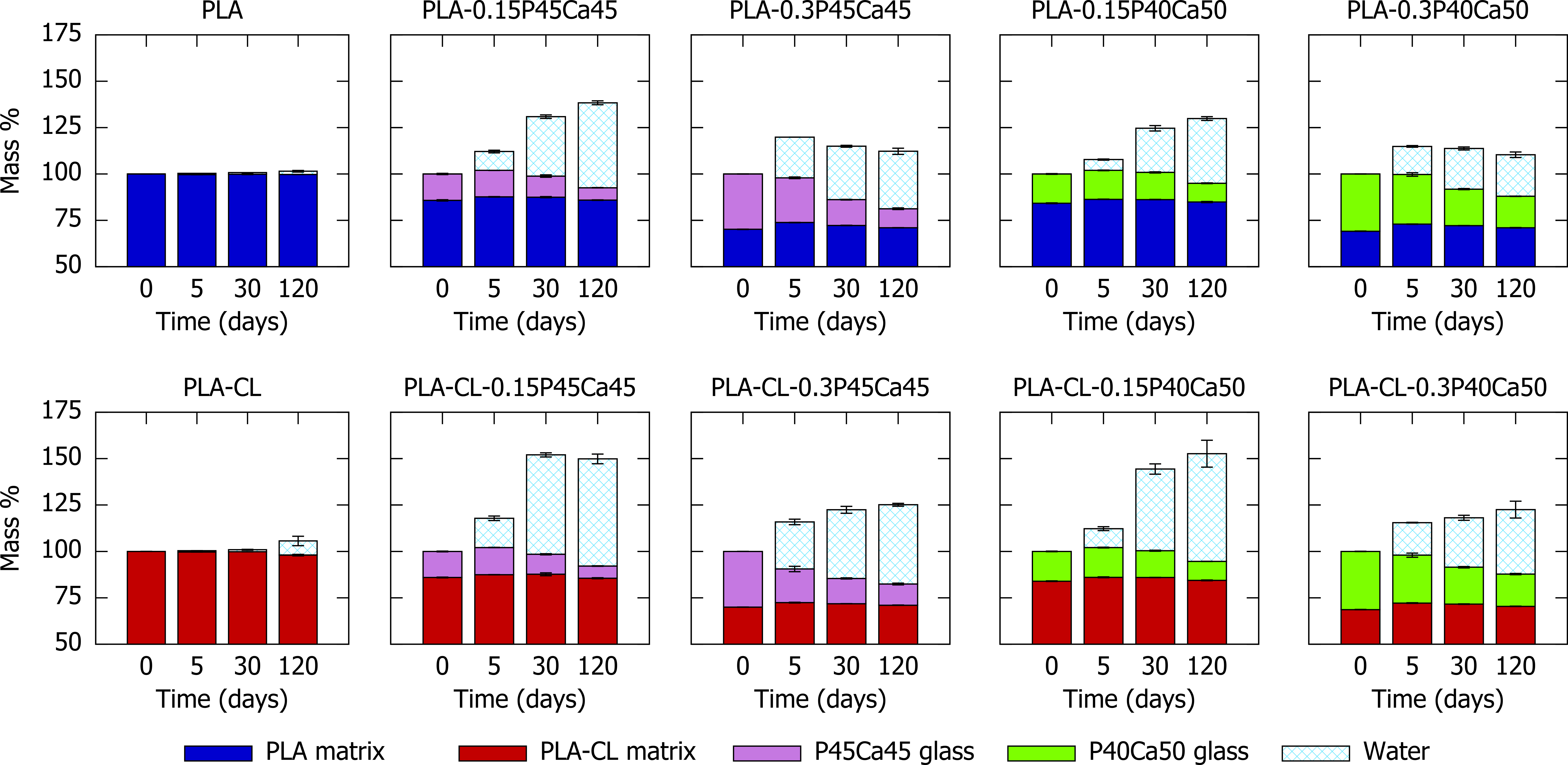}
	\caption{Stacked bar chart with results of drying and ashing tests, showing how component masses change during degradation in PBS at 37\textcelsius. Shown are the water mass, glass mass, and polymer mass percentages. Note that y-axis begins at 50\% as the polymer matrix is always $\geq$ 70\%. All y-axes are identical.}
	\label{fig:CompBars}
\end{figure}

Wet mass measurements can be difficult to interpret due to the overlap of multiple different effects. The drying and ashing procedures allowed these effects to be separated into changes in water, glass, and polymer mass, as shown in Fig. \ref{fig:CompBars}.

Gradual water absorption was seen for all polymer and composite samples, slowing over time. High glass content accelerated water absorption initially (up to 5 days), which can be attributed to the increased glass/polymer interfacial area providing additional area for wicking \cite{Ahmed2011}. After 30 days or more composites with 15wt.\% glass content were exceeded the water absorption of the 30wt.\% glass composites, due in part to the greater proportion of the polymer component, suggesting that although the interface promotes water absorption, the water is stored within the polymer. Greater hydrophilicity of the components also increases water absorption; composites containing a PLA-CL matrix or P45Ca45 glass consistently showed increased water absorption compared with equivalent composites containing a PLA matrix or P40Ca50 glass. Changes in the glass mass were also observed, as the composites absorbed water leading to dissolution of the glass. As expected, the faster dissolving P45Ca45 glass showed faster mass loss, while the polymer matrix composition (PLA or PLA-CL) did not appear to have an impact on glass dissolution. The amount of glass present altered the glass dissolution behaviour, where a delayed start to glass mass loss was observed for composites with lower glass content (minimal mass loss after 5 days), compared with the composites with higher glass content. The slower initial water absorption of the 15wt.\% glass composites is one explanation for this, restricting the water available for dissolution of glass. No change in the polymer mass for PLA samples was observed, while PLA-CL samples displayed a small mass loss as expected (1-2wt.\% of total after 120 days). Within the initial 5 days, these results showed an increase in the polymer mass for all composites, however this is likely to be an artefact of the drying and ashing technique due to incomplete drying.

\subsubsection{Polymer structure}

\begin{figure}[htb]
	\centering
	\includegraphics[width=12cm]{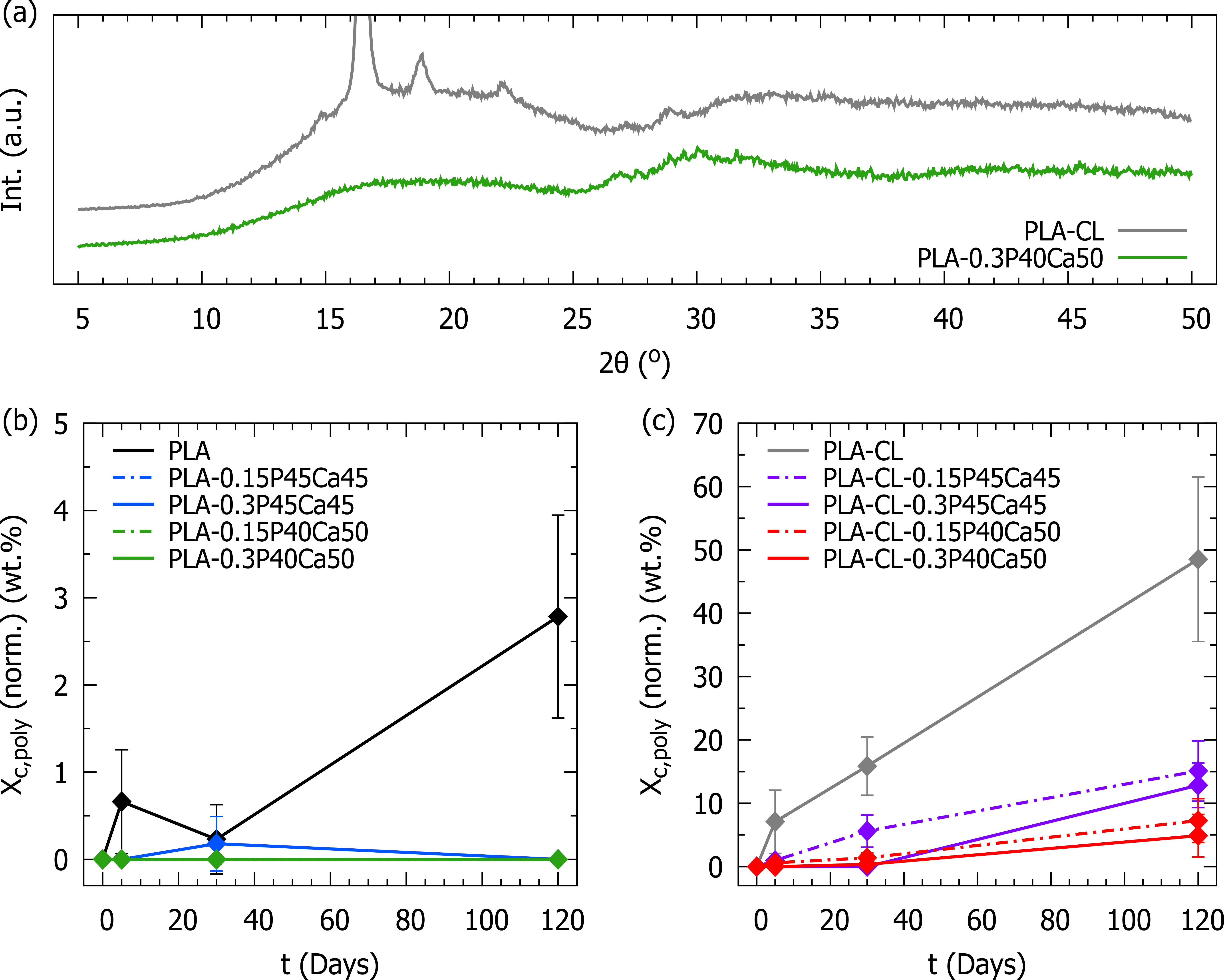}
	\caption{XRD polymer crystallinity (normalised to nominal polymer content) of composites after degradation in PBS at 37\textcelsius. Shown are composites with a PLLA matrix (b) and 90PLLA:10PLCL(70:30)-PEG matrix (c), error bars denote standard deviation for n = 3 samples. Note the different y-axis scale for (b) and (c). (a) shows example XRD patterns after 120 days degradation for samples where significant crystallisation has or has not occurred (PLA-CL and PLA-0.3P40Ca50 respectively).}
	\label{fig:PolyXRD}
\end{figure}

\begin{figure}[htb]
	\centering
	\includegraphics[width=12cm]{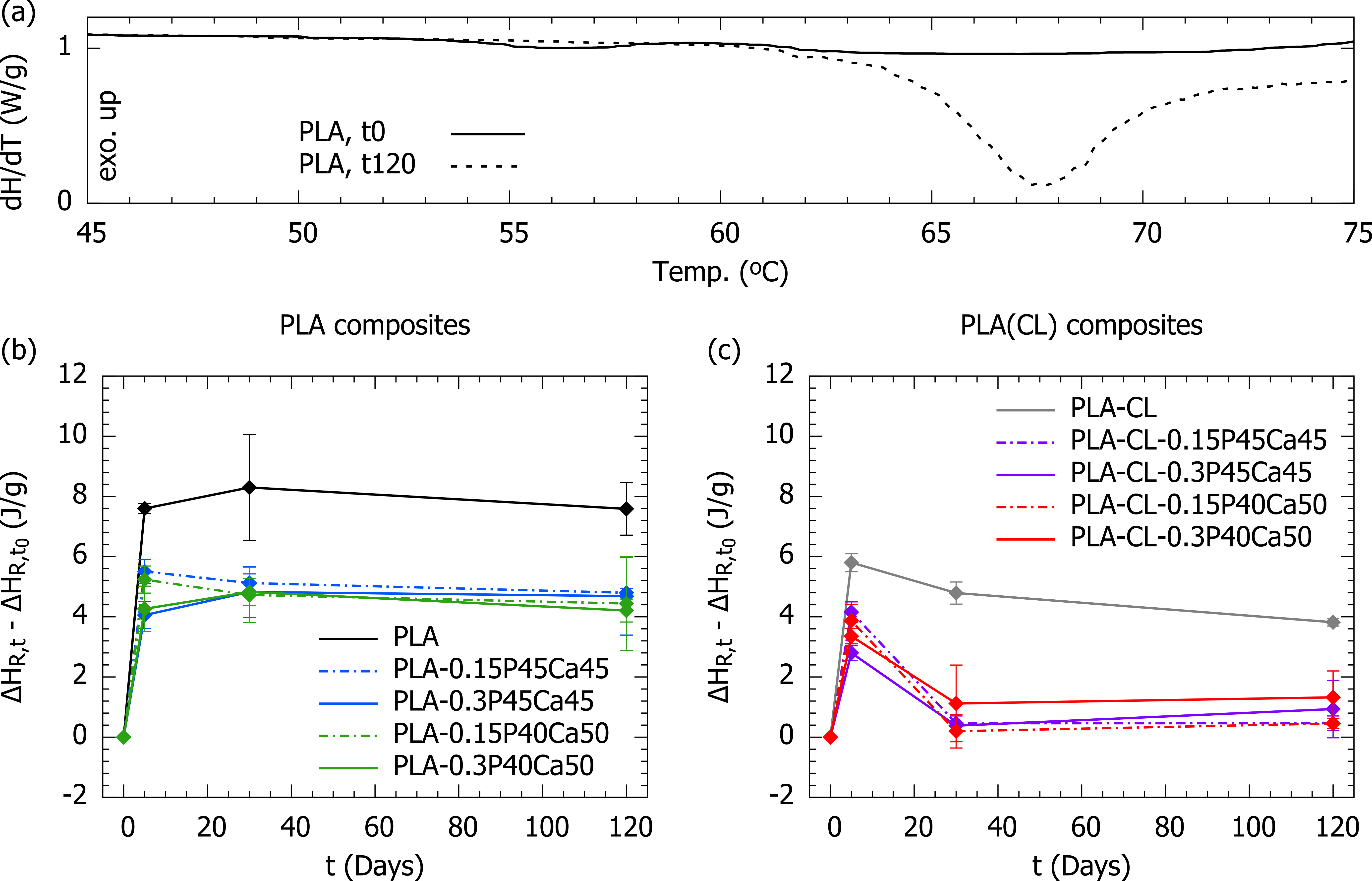}
	\caption{Change in the enthalpy relaxation peak $\Delta H_{R}$ for dry composites measured by DSC between the initial value, and after 5, 30, and 120 days degradation in PBS at 37\textcelsius. Shown are composites with a PLLA matrix (b) and 90PLLA:10PLCL(70:30)-PEG matrix (c), error bars denote standard deviation for n = 3 samples. (a) shows an example of the endothermic peak for PLA before and after 120 days degradation.}
	\label{fig:CompDSC}
\end{figure}

Polymer matrix crystallisation ($\alpha$-PLLA unit cell \cite{Tashiro2017}) was observed during degradation, as shown in Fig. \ref{fig:PolyXRD}. Significantly more crystallisation occurred in the PLA-CL sample than the PLA due to the faster degradation of the PLCL-PEG copolymer and greater mobility of the resulting short chain polymers \cite{Oosterbeek2019}. The addition of phosphate glass to the composite resulted in suppression of crystal formation within the polymer matrix, preventing polymer crystallisation in PLA-based composites, and limiting it to \textless 20\% in PLA-CL-based composites. An initial increase in the enthalpy relaxation ($\Delta H_{R,t} - \Delta H_{R,t_{0}}$) during degradation was observed for all compositions (Fig. \ref{fig:CompDSC}), as a result of increased mobility at the degradation temperature (37\textcelsius ) allowing structural relaxation and densification. This was greater for pure PLA than the PLA-CL blend due to unfavourable interactions between dissimilar polymers \cite{Oosterbeek2019}, while glass addition reduced the extent of enthalpy relaxation in all cases.

\subsubsection{Composite structure}

\begin{figure}[htb]
	\centering
	\includegraphics[width=9cm]{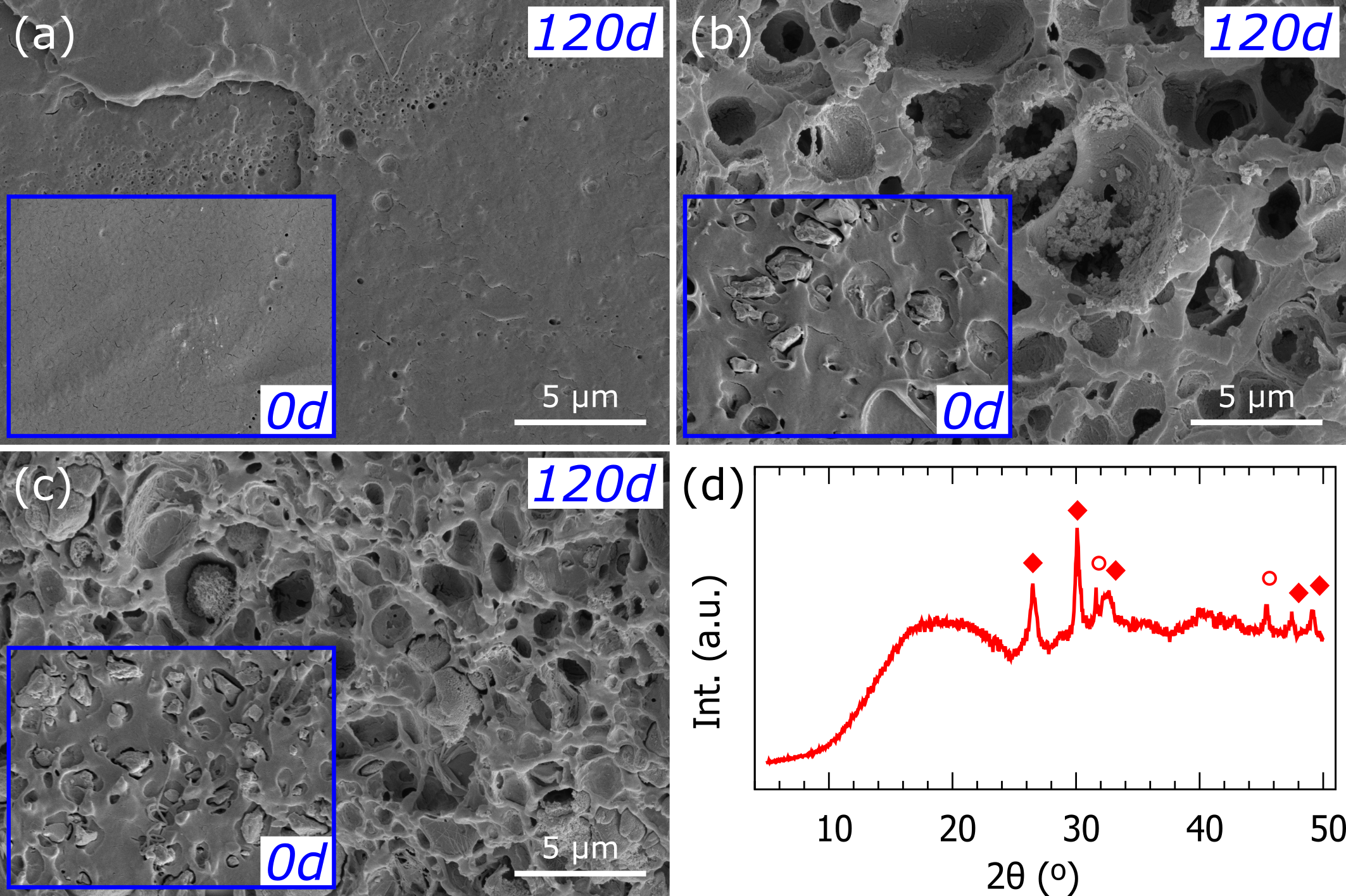}
	\caption{SEM images showing the microstructure of (a) PLA, (b) PLA-0.15P45Ca45, and (c) PLA-0.3P45Ca45 after 120 days degradation, with inset images showing the microstructure before degradation. All SEM images are shown at the same scale. (d) shows an example XRD pattern after 120 days degradation for PLA-0.15P40Ca50 demonstrating inorganic crystallisation, with NaCl (\textopenbullet) and Ca\textsubscript{2}P\textsubscript{2}O\textsubscript{7}.4H\textsubscript{2}O ($\blacklozenge$) phases.}
	\label{fig:CompSEM}
\end{figure}

As a result of glass dissolution and subsequent formation and deposition of conversion layer species \cite{Oosterbeek2021}, inorganic NaCl and Ca\textsubscript{2}P\textsubscript{2}O\textsubscript{7}.4H\textsubscript{2}O crystallites were observed in all composite samples by XRD, reaching a maximum of 3 - 18wt.\% after 120 days degradation (Fig \ref{fig:CompSEM}d). These were also observed with SEM (Fig. \ref{fig:CompSEM}), where originally smooth glass particles showed a roughened surface with crystallite nodules, demonstrating crystal growth or deposition. Polymer only samples show minor morphological changes by SEM, with an increase in small voids seen. The addition of phosphate glass increases this void formation, with 15wt.\% glass having a stronger effect than 30wt.\% glass. Care must be taken in interpreting these SEM images however, because of the potential effect of dehydration on the microstructure. 

\subsection{Evolution of mechanical properties}

\begin{figure}[h!]
	\centering
	\includegraphics[width=12cm]{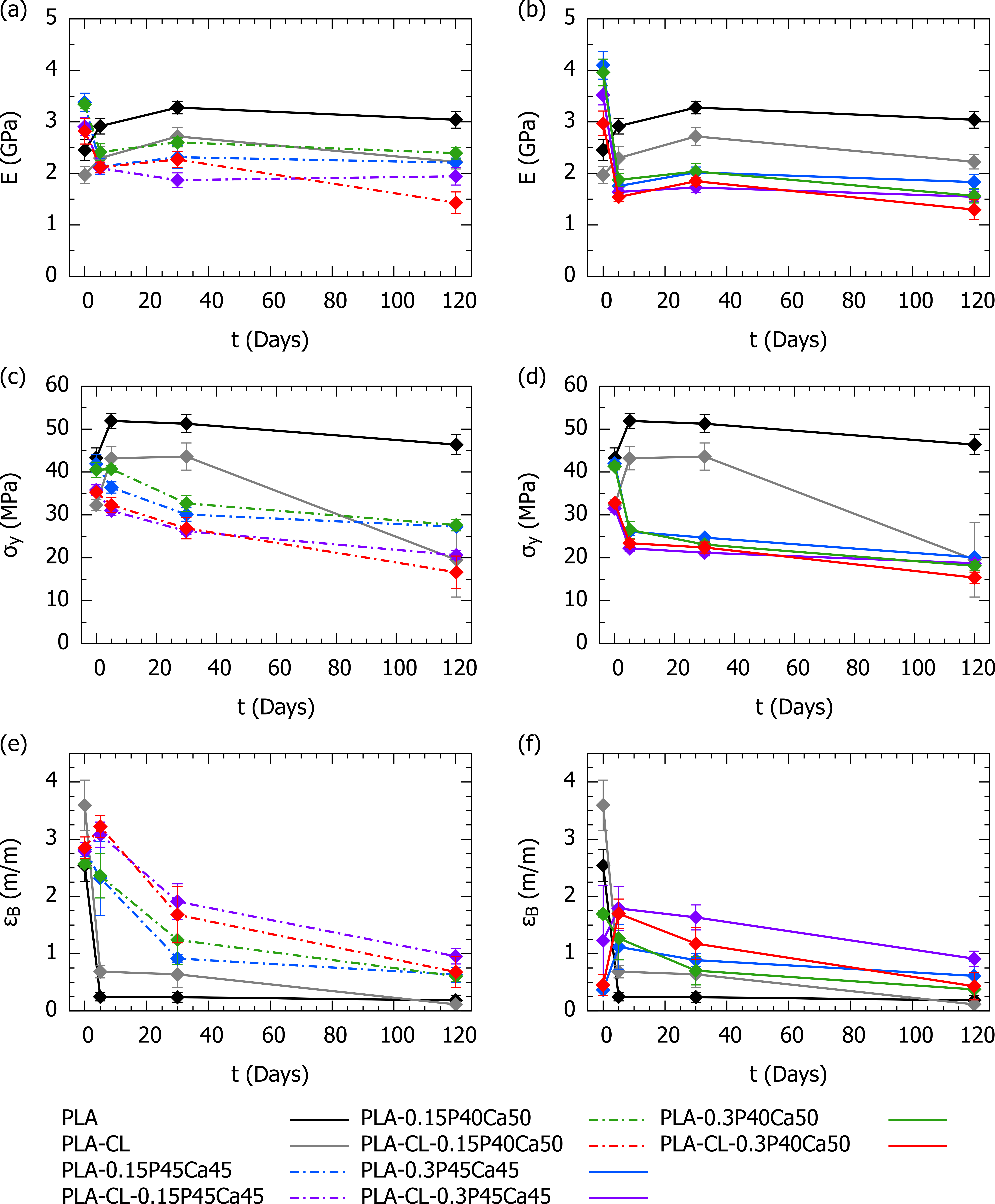}
	\caption{Tensile mechanical properties for polymer-glass composites in 37\textcelsius\ water, measured after 0, 5, 30, and 120 days degradation in PBS at 37\textcelsius. Graphs show elastic modulus $E$ for composites with 15 wt.\% glass (a) and 30 wt.\% glass (b), yield stress $\sigma_{y}$ for composites with 15 wt.\% glass (c) and 30 wt.\% glass (d), and fracture strain $\epsilon_{B}$ for composites with 15 wt.\% glass (e) and 30 wt.\% glass (f).}
	\label{fig:Mech_timeseries}
\end{figure}

As degradation progresses, unfilled polymer samples (PLA and PLA-CL) underwent an increase in elastic modulus until 30 days, followed by a gradual decrease (Fig. \ref{fig:Mech_timeseries}), while filled composites showed a significant modulus decrease within the first 5 days, which was greater for composites with higher glass content. A similar set of trends was observed for the yield strength, with an initial increase followed by reduction for unfilled polymers. Composites with 15wt.\% glass showed a more gradual loss of strength, while those with higher glass content again suffered a larger strength decrease. Similar reductions in composite mechanical properties during degradation are seen in the literature, usually attributed to wicking and water absorption \cite{Zhu2020}.

Unfilled polymers experienced a large reduction in elongation at break ($\epsilon_{B}$) within the first 5 days of degradation, with this reduction more severe for the pure PLA polymer. Composites with 15wt.\% glass showed negligible changes within the first 5 days, and experienced a gradual loss of ductility for longer degradation times, but remained more ductile than the unfilled polymers. Composites with 30wt.\% glass, which initially showed lower and more variable ductility than the 15wt.\% glass composites, generally maintained or increased their ductility over the first 5 days, before also showing gradual ductility loss for longer degradation times.

\section{Discussion}

\subsection{Initial structure and properties}

The measured composite moduli closely match the moduli calculated using the Counto model ($R^{2} = 0.90$), indicating that the main factors affecting the modulus before degradation, are the glass loading and the polymer matrix stiffness. Water absorption, along with the increased temperature under simulated body conditions (immersed in 37\textcelsius\ water), reduces the inter-chain bonding within the polymer \cite{Vyavahare2014} leading to the reduced strength and stiffness under simulated body conditions. When accounting for this reduction in polymer properties, composite moduli under simulated body conditions are still in accordance with the Counto model. Other factors such as the glass composition, interfacial bonding, or deviation from the model at high $\phi_{f}$ are not seen to be significant, and the elastic modulus is thus a function of the matrix stiffness, the stiffness of the inorganic phase, and the loading volume fraction. The use of deionised water for these experiments is a less accurate representation of \textit{in vivo} conditions than PBS, however the effect of this on the observed mechanical testing results is expected to be negligible. There is evidence that reverse osmosis effects can cause polymers to absorb less water in saline solutions, however on the short timescale of these experiments (\textless 20 minutes) this difference is expected to be minimal.

Under ambient conditions, the composite yield strength fell between the bounds defined by Nicolais and Narkis \cite{Nicolais1971} for no adhesion and perfect adhesion, as seen in Fig. \ref{fig:E-yield-t0dry-wet}. This indicated that some stress transfer occurred, however due to imperfect adhesion yield was initiated by separation at the polymer-glass interface, leading to stress concentration in the matrix and subsequent bulk yield. The reduction in yield strength as the filler amount increased appeared more severe for composites with the PLA-CL matrix than the PLA matrix, suggesting that the PLA-phosphate glass interface had superior interfacial adhesion than the PLA-CL-phosphate glass interface, under ambient conditions.

In simulated body conditions (immersed in 37\textcelsius\ water) the yield strength generally followed the upper bound predicted by Nicolais and Narkis \cite{Nicolais1971} shown in Fig. \ref{fig:E-yield-t0dry-wet}b. This could be interpreted as a strengthening of the interfacial adhesion, but is more likely to be a result of the reduced strength of the polymer matrix under non-ambient conditions. This may have led to bulk yield in the polymer matrix before reaching sufficient stress to cause interfacial separation.

Composite ductility variations can mainly be attributed to the behaviour of the polymer matrix, which becomes significantly more ductile in simulated body conditions. In all conditions the effect of particle loading on ductility was similar, where the ductility was relatively unchanged upon addition of 15wt.\% glass, and decreased for the higher 30wt.\% glass loading. This suggests the stress concentrating effect of the filler particles was counteracted by toughening mechanisms (such as interfacial debonding or crack deflection) for the lower 15wt.\% filler loading. For composites with 30wt.\% glass, the increased brittleness may be a result of greater stress concentration, or an effect of particle agglomeration, which is more likely for higher filler loading.

\subsection{Degradation behaviour}

The degradation lifetime can be broadly broken down into two regimes, firstly an initial stage where water absorption is dominant and wet mass increases to a maximum, before the second regime begins, which is dominated by glass dissolution and results in slow mass loss. For composites with 30 wt.\% glass the transition between these regimes occurs at about 5 days, while for composites with 15 wt.\%  glass this takes up to 90 days, as seen in Fig. \ref{fig:Deg_data_all}c-d.

During the first stage, glass composition affects water absorption, with composites containing the more hydrophilic P45Ca45 glass \cite{Uo1998} absorbing more water. The polymer matrix plays an even larger role, with composites where the matrix contains 10 wt.\% of the more hydrophilic PLCL-PEG \cite{Azhari2018} showing significantly higher water absorption. Composite samples show much higher mass increases than the polymers, suggesting that water absorption mainly occurs due to, and along, the interface between the two components \cite{Ahmed2011, Khan2009}. Composites containing more glass will have a greater interfacial area, explaining the faster water absorption experienced by 30 wt.\% glass composites during the initial stage. During this initial stage the Ca\textsuperscript{2+} release and pH change (Fig. \ref{fig:Deg_data_all}a-b) appear to be mainly influenced by the glass dissolution - in spite of the large difference in water absorption caused by addition of PLCL-PEG to the polymer matrix, this only has a small effect on the solution pH and Ca\textsuperscript{2+} concentration. 

In the second regime sample mass change is dominated by glass dissolution. Composites with 15 wt.\% glass show only slight reduction in mass from their peak after 300 days degradation, while composites with 30 wt.\% glass show clear mass loss after a peak at 5 days due to glass dissolution (Fig. \ref{fig:Deg_data_all}c-d). During this regime the glass composition is the dominant factor, with composites containing the faster dissolving P45Ca45 glass losing mass significantly faster than those containing P40Ca50 glass. The addition of PLCL-PEG to the polymer matrix does not have a significant effect, as the mass loss is mainly glass dissolution and the pH reduction caused by glass dissolution has not triggered significant polymer degradation. Within the timescale of the long term degradation experiment, mass loss by polymer degradation of PLA-CL is only just observable after 300 days. 

Despite differences in ionic concentration in the first regime (pH and Ca\textsuperscript{2+}, Fig. \ref{fig:Deg_data_all}a-b), in the second regime these begin to converge as glass dissolution becomes dominant. The solution Ca\textsuperscript{2+} concentration appears to be converging for all different composite compositions (irrespective of glass composition, matrix composition, or glass filler content), suggesting that some equilibrium is being approached between the glass dissolving in the composite, and the solution concentration. This is not a situation that would occur in clinical use of these materials, and represents a limitation of the static degradation test used here.

Increasing the glass filler content from 15 to 30wt.\% has a large effect on the transition time between these two regimes, as the faster water absorption experienced with higher glass content due to the greater interfacial area hastens the onset of glass dissolution. Although increased solution acidity is known to accelerate the rate of PLLA hydrolysis, the modification of solution pH by glass dissolution does not appear to have significantly accelerated the rate of degradation of the polymer component. This suggests that the size of the pH change here was insufficient to trigger catalysis of PLLA hydrolysis \cite{Tsuji2004b}.

\subsection{Structure evolution}

\begin{figure}[htb]
	\centering
	\includegraphics[width=12cm]{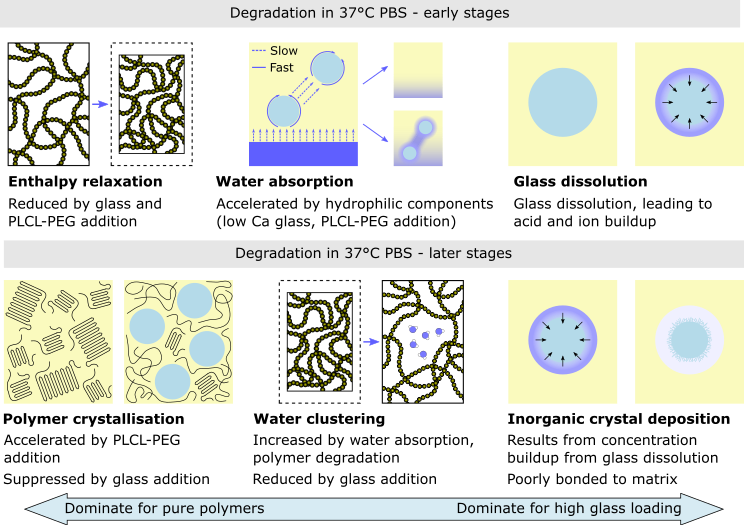}
	\caption{Summary of the mechanisms of structural changes that occur in polymer-glass composites during degradation over different timescales.}
	\label{fig:CompSumm}
\end{figure}

The structural changes occurring within the composites during degradation are summarised in Fig. \ref{fig:CompSumm}. Enthalpy relaxation occurs early on due to increased mobility under simulated body conditions, allowing structural relaxation and densification, and increasing the $T_{g}$ \cite{Pan2007}. Addition of PLCL-PEG to PLLA reduces this enthalpic relaxation due to unfavourable interactions between dissimilar polymer chains \cite{Oosterbeek2019}. The addition of phosphate glass particles also reduces the enthalpic relaxation, due to reduced mobility of the amorphous polymer in the immediate vicinity of a filler particle \cite{Struik1987}. For longer degradation times, crystallisation of polymer chains becomes more significant. This is more pronounced for composites where the matrix contains PLCL-PEG, due to the hydrolysis-induced chain cleavage providing additional mobility for crystallisation \cite{Han2009}. The addition of phosphate glass particles also reduces the polymer crystallisation by reducing the polymer chain mobility in the vicinity of the filler particles.

Water absorption is a key phenomenon determining the structure and properties of these polymer-glass composites during degradation. The increased absorption rate for particulate-filled composites can be attributed to surface interactions of water with the phosphate glass particles \cite{Theocaris1983}. As hydrophilic materials \cite{Uo1998} the phosphate glass will be wetted by absorbed water and contribute to accelerating water transport into the composite by wicking. Absorption is thus increased for higher glass content, and for more hydrophilic components (polymer matrix that includes PLCL-PEG, and lower Ca content in the glass).

At later stages of degradation, water absorption leads to void formation, which was greater for 15wt.\% glass composites than 30wt.\%. Due to the reduced mobility of polymer chains within a composite with greater filler content, the equilibrium void content reduces \cite{Theocaris1983}, explaining the lower water content seen for 30wt.\% glass composites compared with 15wt.\% glass. Although unfilled polymers would theoretically have an even higher equilibrium void, and therefore water content, this is not observed within the timeframe of this experiment due to the faster water absorption seen for composites and the comparatively slow absorption of the unfilled polymers.

Dissolution of glass particles is another key structural change, which dominates for composites with high glass loading and is dependent on water absorption. The reduced interfacial area of 15wt.\% glass composites compared with 30wt.\% glass composites reduces their initial water absorption rate, explaining the initial delay in glass mass loss seen. Mass transport of dissolution products out of the composite is limited, leading to buildup of ions, and although this would lead to acidification, this was apparently insufficient to significantly affect polymer hydrolysis. Concentration buildup does impact glass dissolution behaviour, allowing stabilisation of the conversion layer and triggering the transition from diffusion limited to surface reaction limited glass dissolution. The formation and deposition of conversion layer species was observed in XRD and SEM results, and is responsible for the reduction in glass dissolution rate observed in mass loss measurements \cite{Oosterbeek2021, Ma2017, Ma2018}.

\subsection{Effects on mechanical properties}

Enthalpy relaxation and crystallisation dominate the evolution of unfilled polymer mechanical properties, increasing the elastic modulus and yield strength at the expense of ductility \cite{Pan2007}. The later reduction in the elastic modulus and yield strength of PLA-CL (and to a lesser extent PLA) can be attributed to greater water absorption, and for PLA-CL, hydrolytic degradation. Similar results are seen by Naseem et al. \cite{Naseem2020} who observed a similar initial increase in modulus followed by slow reduction for PLLA, as well as work by Oosterbeek et al. \cite{Oosterbeek2019}.

Changes in composite mechanical properties are a combination of various effects involving the polymer matrix, glass particles, and water absorption. The decrease in elastic modulus is faster and more severe for composites with 30wt.\% glass than 15wt.\%, suggesting initial water absorption is the dominant effect, and causing a physical separation between the glass and polymer matrix (as observed in SEM images). This results in transfer of stress from the glass filler to the polymer matrix, reducing the elastic modulus and yield strength. The reduction seen is greater than would be expected from simple matrix-filler debonding, and can also be attributed to polymer hydration, increasing chain mobility and reducing the elastic modulus and yield strength \cite{Vyavahare2014}. The enthalpy relaxation and polymer crystallisation seen for composites would be expected to counteract this and increase the elastic modulus and yield strength, however these effects are small relative to water absorption and hydration.

Composite ductility also gradually decreases during degradation, except in the initial stages for composites with 30wt.\% glass, where water absorption increases the ductility of those initially brittle composites. The gradual loss of ductility is an effect of absorbed water, causing clustering and void formation, which can act as defects that initiate crack growth \cite{Theocaris1983}. Hydrolytic degradation of the polymer will also play a role in reducing the ductility, and is more significant for composites containing the PLCL-PEG copolymer.

\subsection{Implications for cardiac stents}
The results here, in particular the evolution of mechanical properties over time, give some indication of the suitability of these materials and strategies for bioresorbable cardiac stent applications. The large reduction in elastic modulus and yield strength observed for composites with 30wt.\% glass that occurs within the first 5 days is potentially problematic and could contribute to elastic stent recoil and increase the risk of restenosis \cite{McMahon2018}. By contrast composites of PLLA with 15wt.\% glass initially show comparable strength and superior modulus to pure PLLA and display a more gradual loss of mechanical properties than composites with higher loading, which could be advantageous for a stent and allow gradual transfer of mechanical support to newly healed tissue. Although there is no evidence of glass addition accelerating polymer degradation within the timescale of these experiments, the addition of the PLCL-PEG copolymer to the polymer matrix does accelerate degradation compared with PLLA, allowing the stent to degrade within a more suitable timeframe \cite{McMahon2018, Mukherjee2017}.

The use of Ca-containing bioceramic fillers may appear to pose a risk of arterial calcification or fibrin polymerisation \cite{Weisel2013}, however when the small amount of material, extended timescale of release, and relatively large volume of blood flow through coronary arteries are considered \cite{Ramanathan2005}, the increase in blood Ca levels is negligible compared with the baseline concentration. There is also evidence that the presence of extracellular pyrophosphate ions (P\textsubscript{2}O\textsubscript{7}\textsuperscript{4-}) plays a major role in preventing vascular calcification. Synthesis of calcification inhibitors is the main mechanism by which cells can prevent vascular calcification (i.e. formation of hydroxyapatite). Pyrophosphate ions, one product of phosphate glass dissolution \cite{AbouNeel2008a, Ma2018}, are also one of the main inhibitors of vascular calcification and act by binding directly to nascent hydroxyapatite crystals, preventing further crystal growth \cite{Villa-Bellosta2017}. Again however, the level of this ion release above the baseline may be too small to have an appreciable effect.


\FloatBarrier
\section{Conclusions}

Degradation testing of polymer-glass composites based on PLLA, PLCL-PEG and P\textsubscript{2}O\textsubscript{5}-CaO-Na\textsubscript{2}O glasses demonstrated a two-stage mechanism, where water absorption dominates initially, while later stages are dominated by glass dissolution. The resulting pH reduction did not accelerate polymer degradation within the timescale of these experiments ($\sim$1 year).

For unfilled polymers, chain rearrangements including enthalpy relaxation and polymer crystallisation are critical, while the presence of glass particles suppresses these processes. The glass-polymer interface accelerates water absorption, leading to significant water uptake and cluster formation. This leads to dissolution of the glass component, which dominates for high glass loading, leading to debonding of the glass particles from the matrix.

These structural changes have significant effects on the mechanical properties of the composites during degradation, with unfilled polymers becoming stronger and stiffer, but less ductile. The evolution of mechanical properties is more favourable for cardiac stent applications for composites with 15wt.\% glass, which demonstrate a more gradual loss of mechanical properties during degradation. This could allow gradual transfer of loading to newly healed tissue, and reduce the risk of catastrophic brittle device fracture.

These results provide valuable new understanding of the structural changes that occur during degradation of polymer-glass composites, how these interact with each other, and how the composition of the composite affects which of these mechanisms are dominant. This allows the evolution of mechanical properties during degradation to be understood, which is crucial to designing effective bioresorbable cardiac stent devices.

\section*{Acknowledgements}
The authors thank Lucideon Ltd. for providing materials and financial support, and Ashland Specialties Ireland Ltd. for providing materials. The authors would also like to thank  Mr Wayne Skelton-Hough, Dr Kyung-Ah Kwon, Mr Andrew Rayment, Mr Robert Cornell, and Mr Ian Campbell for their technical support and helpful discussions. RNO would also like to thank the Woolf Fisher Trust and the Cambridge Trust, for provision of a PhD scholarship. Original data for this paper can be found at \url{https://doi.org/10.17863/CAM.63540}.

\bibliographystyle{elsarticle-num}

\bibliography{Refs}

\vspace{5cm}

\begin{flushleft}
	
	Published journal article:\\
	\url{https://doi.org/10.1016/j.compscitech.2021.109194}\\
	\vspace{0.5cm}
	Copyright \textcopyright\ \href{https://creativecommons.org/licenses/by-nc-nd/2.0/uk/}{CC-BY-NC-ND}\\
	\vspace{0.2cm}
	\includegraphics[width=0.2\linewidth]{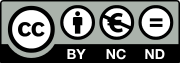}	
	
\end{flushleft}

\end{document}